\documentclass[letterpaper]{ptephy}
\usepackage{amsmath} 
\usepackage{amsthm} 
\newcommand{\bs}{\boldsymbol}
\newcommand{\la}{\langle}
\newcommand{\ra}{\rangle}

\newcommand{\beq}{\begin{eqnarray}}
\newcommand{\eeq}{\end{eqnarray}}

\newcommand{\bfl}{\begin{flushleft}}
\newcommand{\efl}{\end{flushleft}}
\begin{document}

\title{Radiative $\pi^{\pm} \gamma$ transitions 
of excited light-quark mesons 
in the covariant oscillator quark model}

\author{\name{Tomohito Maeda}{1}, 
\name{Kenji Yamada}{1},
\name{Masuho Oda}{2} 
and \name{Shin Ishida}{3}\thanks{Senior Research Fellow}
}
\address{
\affil{1}{
Department of Science and Manufacturing Technology, 
Junior College Funabashi Campus, 
Nihon University, Funabashi 274-8501, Japan
}
\affil{2}{School of Science and Engineering, 
Kokushikan University, Tokyo 154-8515, Japan
}
\affil{3}{
Research Institute of Science and Technology, 
College of Science and Technology, 
Nihon University, Tokyo 101-8308, Japan
}
}
%%%%%%%%%%%%%%%%%%%%%%%%%%%%%%%%%
\begin{abstract}%
The COMPASS collaboration, as a part of their hadron 
spectroscopy program, 
is going to measure the radiative decay widths of light-quark 
mesons via the Primakoff reactions. 
%\cite{COMPASS:2011,COMPASS:2012-1,COMPASS:2012-2}.
In this letter we study the photon couplings of light-quark 
$q\bar{q}$ states in the covariant oscillator quark model 
and evaluate the transition rates for $\{ \rho (770), 
b_{1}(1235), a_{1}(1260), a_{2}(1320), \pi_{2}(1670), 
\rho_{3}(1690), \rho (1700) \}^{\pm} \to \pi^{\pm} \gamma$, 
which are expected to be measured by the COMPASS. 
Such photon couplings could be useful not only for understanding 
the internal structures of observed light-quark mesons and 
their quark-model classification but also for the ongoing 
experimental studies by COMPASS.
\end{abstract}

\subjectindex{D32, B69}

%parindent 0pt

\maketitle
%%%%%%%%%%%%%%%%%%%%%%%%%%%%%%%%%
\section{Introduction} 
The COMPASS is a fixed-target experiment at the CERN SPS 
for investigating the structure and spectrum of hadrons. 
Concerning their hadron spectroscopy program, particular 
attention is paid to light-quark meson systems and one of 
them, the Primakoff production measurements, 
from which the radiative widths of light-quark mesons 
could be obtained, are going on \cite{COMPASS:2011,
COMPASS:2012-1,COMPASS:2012-2}. 

Direct observation of the radiative decays, 
such as $X^{\pm} \to \pi^{\pm}\gamma$, is often 
difficult to carry out because of their very small rates, 
while their inverse reactions, called Primakoff 
reactions \cite{Primakoff:1951, Pomeranchuk-Shmushkevich:1961}, 
which are described as the scattering of a pion in the Coulomb 
field of atomic nuclei $(A,Z)$ 
\beq
\pi^{\pm} + (A, Z) 
\to \pi^{\pm} \gamma^{*}+ (A, Z) \to X^{\pm} + (A, Z), \ 
\label{Eq:0}
\eeq 
are relatively easy of access. Since the cross section for 
reaction (\ref{Eq:0}) at very low-$q^2$ regions 
($q$ being the four momentum transfer) is proportional to 
the radiative decay width 
$\Gamma (X^{\pm} \to \pi^{\pm} + \gamma)$, 
it is possible to determine them by measuring 
the Coulomb contribution of the absolute cross section 
\cite{Zielinski:1987}. 
While a number of new experimental measurements 
were performed in recent years \cite{PDG:2012}, 
it should be noted that some earlier experiments, 
such as Ref.~\cite{Zielinski:1984,Collick:1984}, 
have been cited as the latest data. Thus there is 
no doubt that the high statistics data from COMPASS  
will play an important role in the progress of 
light meson spectroscopy. 

Theoretically it has been well known that radiative transitions 
probe the internal structures of hadrons and hence it offers 
a useful tool to investigate their nature. In the transitions of 
excited states of light-quark mesons, although the final mesons 
have large kinetic energy at the rest frame of initial ones, 
such an effect is neglected in the conventional treatment of 
naive nonrelativistic quark models (NRQM). 
In addition, respecting the actual situation that 
all physical observations are made through not quarks but hadrons, 
the relativistic treatment for the center-of-mass (CM) motion 
of hadrons is absolutely necessary. 

The covariant oscillator quark model (COQM) is 
one of the possible covariant extension of NRQM, 
retaining the various success principally restricted to the static problem. 
The remarkable features of the COQM is 
that hadrons are treated in a manifestly 
covariant way and the conserved effective electromagnetic currents
of hadrons are explicitly given in terms of hadron variables themselves. 
{\footnote{
The COQM has a long history of development.
It had been applied to investigate the radiative decays 
of light-quark meson systems {\cite{Ishida-Yamada-Oda:1989}} and 
heavy quarkonium systems {\cite{Ishida-Morikawa-Oda:1998}}
with considerable success. }

In this work we shall apply the COQM 
to analyze the one photon couplings in the transitions 
\{$\rho(770)({}^{3}S_{1}), ~b_{1}(1235)({}^{1}P_{1}), ~a_{1}(1260) ({}^{3}P_{1}), 
~a_{2}(1320)({}^{3}P_{2}), ~\pi_{2}(1670)({}^{1}D_{2}), 
~\rho_{3}(1690)$\\$({}^{3}D_{3}), ~\rho (1700)({}^{3}D_{1})$\}$^{\pm}$ $\to 
\pi({}^{1}S_{0})^{\pm}\gamma$. 
The obtained radiative decay widths are compared with experiment 
and also other quark model predictions. 
The rates for excited $D$-wave states are newly 
predicted and these results could provide a useful clue to 
observe the radiative decays of excited light-quark mesons 
at COMPASS. 
%%%%%%%%%%%%%%%%%%%%%%%%%%%%%%%%%%%%%%%%%%
\section{Basic framework of the COQM}
Let us briefly summarize the framework of 
the COQM relevant to the present application.  In the COQM 
the wave function (WF) of $u\bar{d}$ mesons\footnote{
In the following we restrict ourselves to the case of $u\bar{d}$ mesons. }
is given by the bilocal bispinor field 
$\Psi(x_{1\mu},x_{2\mu})_{\alpha}{}^{\beta}
=\Psi(X_{\mu},x_{\mu})_{\alpha}{}^{\beta}
$, where $\alpha$, $\beta$ denote the Dirac spinor indices
of respective constituents and $x_{1}^{}$, $x_{2}^{}$ 
represent their space-time coordinates, which are 
related with the CM and relative coordinates given by 
$X_{\mu}=(x_{1\mu}+x_{2\mu})/2$ and 
$x_{\mu}=x_{1\mu}-x_{2\mu}$, respectively. 
The WF is assumed to satisfy the
following basic equation {\cite{Yukawa:1953}} 
\beq
\label{Eq:1}
\left(\sum_{i=1}^{2}\frac{-1}{2m}\frac{\partial^{2}}{\partial x^{2}_{i\mu}}
+\frac{K}{2}(x^{}_{1\mu}-x^{}_{2\mu})_{}^2 \right) 
\Psi(x_{1}^{}, x_{2}^{})_{\alpha}{}^{\beta}=0, 
\eeq
which is equivalently rewritten as 
\beq
\label{Eq:2}
\left(
-\frac{\partial^2}{\partial X_{\mu}^{2}}+{\mathcal{M}}^2( x, 
\frac{\partial}{\partial x_{}}) \right)
\Psi(X, x)_{\alpha}{}^{\beta}=0, \ \ 
{\mathcal{ M}}^2 =
d \left( -\frac{1}{2\mu}\frac{\partial^2}
{\partial x_{\mu}^2}+\frac{K}{2}x_{\mu}^2\right), 
\eeq
where ${\mathcal{ M}}^2$ represents the spin-independent 
squared-mass operator in the pure confining force limit, 
$d=4m$, $\mu=m/2$ ($m$ being the effective quark mass) 
and $K$ is the spring constant. 
In order to freeze the redundant freedom 
of relative time, we adopt the definite-metric type of
subsidiary condition for the four dimensional harmonic oscillator (HO),
{\footnote{The WF satisfying this condition 
is normalizable and gives the desirable asymptotic 
behavior of electromagnetic form factors of hadrons {\cite{Takabayashi}.}}
} leading to the eigenvalue solutions of 
the squared-mass operator as 
\beq
\label{Eq:4}
M_{N}{}^2=M_{0}{}^2+N\Omega, \ \ N=2N_{r}+L,
\eeq
where $N_{r}$ and $L$ are the radial and 
orbital quantum numbers respectively and 
$\Omega$ is given by 
$\Omega=d\sqrt{K/\mu}=\sqrt{32mK}$. 
The relation (\ref{Eq:4}) is 
in accord with the well-known linear rising Regge trajectory 
concerning the squared mass spectra, which is particularly 
evident in light-quark hadron sectors. 

The WF describing mesons with the CM four momentum $P_{\mu}$ 
can be written as 
\beq
\label{Eq:5}
\Psi (x_{1}, x_{2})_{\alpha}^{(\pm)}{}^{\beta} 
=\frac{1}{\sqrt{2P_{0}}} e^{\pm P_{\mu}X_{\mu}} 
\Phi (v, x)_{\alpha}{}^{\beta (\pm)}, 
\eeq
where $v_{\mu}=P_{\mu}/M$ is the four velocity ($M$ being the meson masses).
The internal WF $\Phi (v, x)$ is taken as a form of the
``$LS$-coupling'' product in an analogous fashion to NRQM, 
\beq
\label{Eq:6}
\Phi (v, x)^{(\pm)}_{\alpha}{}^{\beta} =f(v, x)^{(nL)}_{\mu\nu\cdots} \otimes
\left( W^{(\pm)}_{\alpha}{}^{\beta}(v)\right)_{\mu\nu\cdots}, 
\eeq
where $f(v, x)^{(nL)}_{\mu\nu\cdots}$ with $n=N+1$ are
the definite-metric-type wave functions of the four dimensional HO 
for the space-time part and $W^{(\pm)}_{\alpha}{}^{\beta}(v)$ 
are the Bargmann-Wigner spinor 
functions{\footnote{They are defined by the direct tensor-product 
of respective constituent 
Dirac spinors with the four velocity of mesons as 
$W_{\alpha}^{(+)\beta}(v) \sim u_{\alpha}(v)\bar{v}^{\beta}(v)$, 
$W_{\alpha}^{(-)\beta}(v)\sim v_{\alpha}(v)\bar{u}^{\beta}(v)$, 
}} for the spin part. 
Decomposing the $W^{(+)}_{\alpha}{}^{\beta}(v)$ 
into irreducible components with the definite $J^{PC}$, 
the detailed expressions for meson states relevant to 
the present study are obtained as follows: 
\begin{subequations}
\beq
\label{Eq:7}
\Phi (v, x)^{(+)}=f^{(1S)}(v, x)W^{(+)}(v)=f_{0}(v,x)
\left( \frac{1+iv_{\rho}\gamma_{\rho}}{2\sqrt{2}}\left(
-\gamma_{5}+i\gamma_{\mu} \epsilon_{\mu}^{}(P)
\right) \right)
\eeq
for the $S$-wave states, 
\beq
\label{Eq:8}
\lefteqn{
\Phi (v, x)^{(+)}=f^{(1P)}(v, x)_{\nu}  W^{(+)}(v)_{\nu}}
\nonumber\\
&&=
\sqrt{2\beta^2} x_{\nu} f_{0}(v, x)
\left( \frac{1+i{v}_{\rho}\gamma_{\rho}}{2\sqrt{2}}
\left(-\gamma_{5} \epsilon^{}_{\nu}(P)+i\gamma_{\mu} \epsilon
_{\mu\nu}(P)\right)\right) 
\eeq
for the $P$-wave states and 
\beq
\label{Eq:8e}
\lefteqn{\Phi (v, x)^{(+)}
=f^{(1D)}(v, x)_{\nu\lambda}  W_{}^{(+)}(v)_{\nu\lambda}}\nonumber\\
&&=2\beta^2 x_{\nu} x_{\lambda} f_{0}(v, x)
\left( \frac{1+iv_{\rho}\gamma_{\rho}}{2\sqrt{2}}
\left(-\gamma_{5} \epsilon_{\nu\lambda}(P)+i\gamma_{\mu} 
\epsilon_{\mu\nu\lambda}(P)\right)\right) 
\eeq
\end{subequations}
for the $D$-wave states, where $f_{0}(v,x)$ is the ground-state
WF of the four dimensional HO given by 
\beq
f_{0}(v,x)=\left(\frac{\beta^2}{\pi}\right) \exp 
\left(-\frac{\beta^2}{2}\left( x_{\sigma}^2+
2(v_{\sigma}x_{\sigma})^2\right)\right)
\label{Eq:12}
\eeq
with the parameter $\beta^2=\sqrt{\mu K}$ 
and $\epsilon_{\mu\cdots}$ are the
polarization tensors for respective meson states. 
%%%%%%%%%%%%%%%%%%%%%%%%%%%%%%%%
\section{Electromagnetic Meson Current in the COQM} 

Next, we introduce the single photon couplings of 
$q\bar{q}$ meson systems. The relevant 
decay amplitude is described by 
\beq
\label{Eq:9}
\int d^4 X \la f | {\mathcal{H}}_{\rm int.} |i \ra 
= -\int d^4 X \la f |J_{\mu}(X) A_{\mu}(X) |i \ra 
=\sqrt{\frac{1}{8P_{I0}P_{F0}q_{0}}}\delta^{4}(P_{I\mu}-P_{F\mu}-q_{\mu})
\mathcal{M}_{fi}, 
\eeq
where $P_{I\mu}$, $P_{F\mu}$ and $q_{\mu}(=P_{I\mu}-P_{F\mu})$ are 
the four momenta of initial and final 
mesons and emitted photon, respectively. 
In the COQM, as is the case in NRQM, we consider that the
one-photon emission proceeds through a single quark transition 
in which the respective constituents couple with a photon. 
In order to obtain the electromagnetic current of mesons, 
we perform the minimal substitutions 
\cite{Feynman-Kislinger-Ravndal, Lipes:1972}
$
%\label{Eq:10}
{\partial}/{\partial x_{i\mu}}\to 
{\partial}/{\partial x_{i\mu}}-ieQ_{i}A_{\mu}(x_{i}) \ \ (i=1,2)
$
%\eeq
in Eq.(\ref{Eq:1}) 
and then obtain
%{\footnote{
%Here the $j^{(d)}_{\mu}$ can also be obtained 
%by the replacement $u\leftrightarrow d$ 
%to Eq.~(\ref{Eq:11}). }} 
\beq
\label{Eq:11}
j_{i\mu} (x_{1},x_{2})&=&j^{(\rm convec)}_{i\mu} (x_{1},x_{2})+j^{(\rm spin)}_{i\mu} (x_{1},x_{2}) 
\eeq
with
\beq
j^{(\rm convec)}_{i\mu} (x_{1},x_{2})
&=&-\frac{ideQ_{i}}{2m} \la \bar{\Psi}(x_{1},x_{2})
\left(\frac{\overleftrightarrow{\partial}}{\partial x^{}_{i\mu}} 
\right)
\Psi (x_{1},x_{2})\ra, \\
%-\frac{\overleftarrow{\partial}}{\partial x^{(u)}_{\mu}}
%\nonumber\\
%\ \ \ \ \ \ 
j^{(\rm spin)}_{i\mu} (x_{1},x_{2})
&=&-\frac{ideQ_{i}}{2m} \la \bar{\Psi}(x_{1},x_{2})
\left(ig_{M}^{(i)}\sigma_{\mu\nu}(\frac{\overrightarrow{\partial}}{\partial x^{}_{i\nu}}+\frac{\overleftarrow{\partial}}{\partial x^{}_{i\nu}}) \right) 
\Psi (x_{1},x_{2})\ra,
%\nonumber\\
\eeq
where $\la \cdots \ra$ means taking trace concerning 
the Dirac indices, $\bar{\Psi}\equiv -\gamma_{4} \Psi^{\dagger} \gamma_{4}$,
$Q_{i}$ represent quark charges in units of $e$ 
($Q_{1}=Q_{u}=2/3$ and $Q_{2}=Q_{d}=-1/3$ for the present application) 
and $g_{M}^{(1)}=g_{M}^{(2)}\equiv g_{M}$ are the parameters 
concerning the anomalous magnetic moment of quarks. 
It is worth mentioning that there exist two kinds of current, 
$j^{(\rm convec)}$ and $j^{(\rm spin)}$ 
(denoting the convection and spin currents respectively), 
which are conserved independently. 

Substituting Eq.(\ref{Eq:5}) and 
%\beq
%\label{Eq:13}
$A_{\mu}(x_{i}) =(1/\sqrt{2q_{0}})
e_{\mu}^{*}(q) 
e^{-iq_{\mu} x_{i\mu}} 
$
%\eeq
into 
\beq
\label{Eq:12}
-\int d^4 x_{1} \int d^{4}x_{2}~\la f|
 j_{1\mu}(x_{1},x_{2})  A_{\mu}(x_{1})|i \ra
+(1\leftrightarrow 2) ~~~~~~
\eeq
and equating it with the matrix element Eq.(\ref{Eq:9}), 
we obtain a formula to calculate 
the decay amplitudes as 
\beq
\label{Eq:14}
\mathcal{M}_{fi}
&=&-eQ_{1}\int d^4 x 
\la\bar{\Phi}^{(-)}_{F}(v_{F}, x)
\left( P_{I\mu}+P_{F\mu} \right. \nonumber \\
&&- \frac{d}{2m}
i\frac{\stackrel{\leftrightarrow}{\partial}}{\partial x_{\mu}}
+\frac{d}{2m}g_{M}\sigma_{\mu\nu}iq_{\nu} )
\Phi^{(+)}_{I}(v_{I},x) \ra e^{-i\frac{2m}{d}q_{\rho}x_{\rho}}e^{*}_{\mu}(q)
+\left(1\leftrightarrow 2\right), 
\eeq
where $e^{}_{\mu}(q)$ is the polarization vector 
of the photon. 
%In the actual calculation, the second term of Eq.(\ref{Eq:14}) including 
%the derivative of the internal coordinate is 
%replaced by 
%\beq
%\label{Eq:15}
%\langle f|\left( -\frac{d}{2m}i
%\frac{\stackrel{\leftrightarrow}{\partial}}{\partial x_{\mu}}\right)|i \rangle
%= -i\frac{\mu}{m} \langle f|
%[ x_{\mu}, {\mathcal{M}}^2(x,\partial/\partial x)]|i \rangle
%=-i\frac{\mu}{m}(M_{I}^2-M_{F}^2)\langle f|x_{\mu}|i \rangle
%\eeq
%so as to 
%respect the low-energy theorem in the electromagnetic interactions, as well as 
%conventional treatment of the NRQM. 
By using this formula, we can derive 
the covariant expressions of invariant amplitudes for the respective radiative 
transitions summarized in Table \ref{tab:1}. 
%decay amplitudes of respective radiative 
%transitions. In Table \ref{tab:1} we have collected them 
%and summarized the relevant radiative decay widths. 
%%%%%%%%%%%%%%%%%%%%%%%%%%%%%%%%%%%%
\section{Numerical Predictions}

Taking the following values of parameters, 
%appeared in the application of the COQM. \\
\begin{itemize}
\item $M_{0}=0.75$ GeV and $\Omega=1.11$ GeV$^2$ \cite{Oda:1999},
which give $M_{1}=1.29$ GeV and $M_{2}=1.67$ GeV 
\item $g_{M}=0.82$, determined from the experimental width of
$\rho^{\pm} \to \pi^{\pm} \gamma$ 
\end{itemize} 
we calculate numerical values of the radiative decay widths.
The results are shown in Table \ref{tab:3} in comparison with 
experiment and other quark-model predictions. 
From this table we can see that our results are in fairly 
agreement with experimental values. 
%%%%%%%%%%%%%%%%%%%%%%%%%%%%%%%%%%%%%%%%%%%%
\begin{table}[htbp]
\vspace{-1em}
%\begin{center}
\caption{Invariant amplitudes and formulas of decay width in the relevant transitions. 
Here $\epsilon_{\mu\cdots}$ and $|\bs{q}_{\gamma}|$ denote the
polarization tensors for initial mesons and physically emitted photon momentum 
at the rest frame of initial mesons, respectively, $\omega=-v_{I\mu} v_{F\mu}=({M_{I}^2+M_{F}^2})/(2M_{I}M_{F})$, and $F$ is 
defined in the text. }
\begin{tabular}{lll}
\hline
%\multicolumn{3}{l}{${}^{2S+1}L_{J}\left(\epsilon_{\mu\cdots}^{}, M_{I}, P_{I}\right)
%\to {}^{1}S_{0}\left(M_{F}, P_{F}\right)+e^{*}\left(q\right)$}\\
$j^{(\rm spin)}_{\mu}$ Process &
$\mathcal{M}_{fi}
%({}^{2S+1}L_{J}\left(\epsilon_{\rho\cdots}^{}, ~M_{I}, ~P_{I}\right)
%\to {}^{1}S_{0}\left(M_{F},~P_{F}\right)+\gamma(e^{*}_{\mu}, ~\bs{q}_{\gamma}))
$
&\ $\Gamma$\\
\hline
${}^{3}S_{1}\to {}^{1}S_{0} \gamma$ \ &
$-e g_{\rho} \varepsilon_{\mu\nu\rho\alpha}e^{*}_{\mu}(q)q_{\nu}
\epsilon_{\rho}(P_{I}) P_{I\alpha}$
&$\frac{4\alpha}{3} |\bs{q}_{\gamma}|^3 g_{\rho}^2 $
\\
 
${}^{3}P_{J=1,2} \to {}^{1}S_{0} \gamma$ \ &
$ieg_{a}\varepsilon_{\mu\nu\rho\alpha}e^{*}_{\mu}(q)q_{\nu}
\epsilon_{\rho\lambda}(P_{I})q_{\lambda} P_{\alpha} $
&$\frac{\alpha}{2(2J+1)} 
|\bs{q}_{\gamma}|^5 g_{a}^2$\\

${}^{3}D_{J} \to {}^{1}S_{0} \gamma$ \ &
$eg_{\rho_{D}}\varepsilon_{\mu\nu\rho\alpha}q_{\nu}\epsilon_{\rho\sigma\lambda}(P_{I})q_{\sigma}
q_{\lambda}P_{I\alpha}e^{*}_{\mu}(q)$
&$C_{J}\alpha |\bs{q}_{\gamma}|^{7}g^2_{\rho_{D}}$ \\

\hline
$j^{(\rm convec.)}_{\mu}$ Process &
$\mathcal{M}_{fi}$
&\ $\Gamma$\\
\hline

${}^{1}P_{1} \to {}^{1}S_{0} \gamma$ \ &
$i eg_{b_{1}} \epsilon_{\mu}(P_{I})e_{\mu}^{*}(q)$
&$\frac{\alpha}{3M^2_{I}} |\bs{q}_{\gamma}| g^2_{b_{1}}$
\\

${}^{1}D_{2} \to {}^{1}S_{0} \gamma$ \ &
$eg_{\pi_{2}}
(e^{*}_{\mu}(q)\epsilon_{\mu\kappa}(P_{I}) q_{\kappa}
+q_{\kappa} \epsilon_{\kappa\mu}(P_{I}) e^{*}_{\mu}(q))$&
$\frac{\alpha}{10M^2_{I}} |\bs{q}_{\gamma}|^3 g^2_{\pi_{2}}$\\
%&\ \ \ \  \ \ \ \ \ $$
%&
%\\
\hline
{Coupling parameters:}&\\
\multicolumn{3}{l}{$
g_{\rho}= g_{M} (Q_{u}+Q_{d})\left(\frac{1}{2M_{I}}+\frac{1}{2M_{F}}\right)F, \ \ 
g_{b_{1}}=\frac{1}{\sqrt{2\beta^2}} \frac{Q_{u}+Q_{d}}{2}\left(M_I^2-M_F^2\right) \frac{1+\omega}{2}F, \ \ 
$}\\

\multicolumn{3}{l}{$
g_{a}=g_{M}(Q_{u}-Q_{d})
(\frac{1}{2M_{I}}+\frac{1}{2M_{F}}) \frac{1}{\sqrt{2\beta^2}}\frac{M_{I}}{\omega M_{F}}F,\ \ \ 
g_{\pi_{2}}=\frac{1}{2\beta^2}
\frac{Q_{u}-Q_{d}}{4}\left(M_I^2-M_F^2\right)\frac{1+\omega}{2\omega}\frac{{M_{I}}}{{M_{F}}}
F, \ \ \ 
$}\\

\multicolumn{3}{l}{$
g_{\rho_{D}}=g_{M}\frac{Q_{u}+Q_{d}}{2}
\frac{1}{2\beta^2}\frac{M_{I}}{\omega^2 M_{F}}
\left(\frac{1}{2M_{I}}+\frac{1}{2M_{F}}\right) F, \ \ \ \ C_{J}=\{ \frac{4}{105},\frac{1}{15}, 
\frac{1}{45}\} \ {\rm for} \  \{ {}^{3}D_{3}, {}^{3}D_{2}, {}^{3}D_{1} \}
$}\\

\hline
\end{tabular}
\label{tab:1}
%\end{center}
\end{table}
%%%%%%%%%%%%%%%%%%%%%%%%%%%%%%%%%%%%
%%%%%%%%%%%%
\begin{table}[t]
%\begin{center}
\caption{Calculated widths in 
comparison with experiment and other models.  }
\begin{tabular}{llllll}
\hline
&&\multicolumn{4}{c}{$\Gamma$/keV}\\
\cline{3-6}
Process &$|\bs{q}_{\gamma}|$/GeV&This work & Experiment \cite{PDG:2012}&
Ref.~\cite{Godfrey-Isgur:1985}&Ref.~\cite{Rosner:1981}
\\
\hline
\underline{$j^{(\rm spin)}$ Process} &&&\\
$\rho^{\pm}\to \pi^{\pm} \gamma$ &0.375&
68 (input)& 
68 $\pm$ 7&68 (input)
&-
\\

$a_{1}(1260)^{\pm} \to \pi^{\pm} \gamma$&0.608&
278&640$\pm$246~\cite{Zielinski:1984}
&314&(1.0-1.6)$\cdot 10^3$
\\
$a_{2}(1320)^{\pm} \to \pi^{\pm} \gamma$&0.652&
237&287$\pm$ 30
&302(input)&375$\pm$50
\\

$\rho_{3}^{}(1690)^{\pm} \to \pi^{\pm} \gamma$&0.839 &21&
-&-&-\\

$\rho_{}(1700)^{\pm} \to \pi^{\pm} \gamma$&0.854&
14&
-&-&-\\

\underline{$j^{(\rm convec.)}$ Process} &&&\\
$b_{1}^{}(1235)^{\pm} \to \pi^{\pm} \gamma$ &0.607&57.8 (Model A) 
&230$\pm$60~\cite{Collick:1984}
&397&184$\pm$30
\\
&&
116~ (Model B)&
&&
\\

$\pi_{2}(1670)^{\pm} \to \pi^{\pm} \gamma$&0.829&
335~ (Model A)&-
&-&-\\
&&
521~ 
(Model B)
&-&-&-\\

\hline
\end{tabular}
\label{tab:3}
%\end{center}
\end{table}
%%%%%%%%%%%%%%%%%%%%%%%%%%%%
\section{Discussion}
Let us discuss in detail about obtained results. 
At first we would like to remark that 
a relativistic effect of the transition form factor $F$ 
commonly contained in all decay amplitudes, 
plays a important role throughout this work. 
It is given by the overlapping integral of 4-dimensional 
HO functions 
\vspace{-1em}
\beq
F &=&\int d^4 x~ f_{0}(v_{F}, x) f_{0}(v_{I}, x)e^{-i\frac{q_{\mu}x_{\mu}}{2}}=
\frac{1}{\omega}\exp\left(-\frac{1}{16\beta^2}\frac{2v_{I\mu}q_{\mu}v_{F\nu}q_{\nu}}{\omega}\right)
\nonumber\\
%\frac{2M_{I}M_{F}}{M_{I}^2+M_{F}^2}\exp
%\left(-\frac{1}{16\beta^2}\frac{(M^2_{I}-M^2_{F})^2}{M^2_{I} +M^2_{F}}\right)
&\stackrel{\bs{P}_{I}= \bs{0}}{=}& \frac{M_{F}}{(P_{F})_{0}}\exp\left(-\frac{|\bs{q}|^2}{16\beta^2}  
2(1+\frac{|\bs{P_{F}}|}{(P_{F})_{0}}) \right). 
\label{Eq:18}
\eeq
%which corresponds to 
%instead of the familiar 3-dimensional one, 
%\beq
%F^{\rm 3D} =\int d^3 x~ f_{0}(\bs{x}) f_{0}^{}(\bs{x})e^{-i\frac{\bs{q}\cdot \bs{x}}{2}}=
%\exp\left(-\frac{|\bs{q}|^2}{16\beta^2} \right).  \ \ \ (|\bs{q}|=\frac{M_{I}^2-M_{F}^2}{2M_{I}})
%\eeq
%For example, 
%in the transition $a_{2}(1320)^{\pm} \to \pi^{\pm} \gamma$, 
%it leads $F=0.677$, while $F^{\rm 3D}=0.920$. 
%It can be seen clearly when we take the rest frame of 
%initial particle ($\bs{P}_{I}=\bs{0}, \bs{q}=q\hat{z}$)
%\beq
%F =\int d^4 x~ f^{(1S)}(v_{F}, x) f^{(1S)}(v_{I}, x)e^{-i\frac{q_{\mu}x_{\mu}}{2}}{\to} 
%\frac{M_{F}}{(P_{F})_{0}}\exp\left(-\frac{|\bs{q}|^2}{16\beta^2}  2(1+\frac{|\bs{P_{F}}|}{(P_{F})_{0}}) \right),  
%\label{Eq:18}
%\eeq
%where $|\bs{q}|=(M_{I}^2-M^2_{F})/2M_{I}$, leading to the strong damping effect 
%Note that factor 2 in the exponent of Eq.~(\ref{Eq:18}) arises 
%from the relative time degree of freedom in the WF. 
%Accordingly our results yield the ratio 
%Accordingly, in our results, 
Resulting ratio is $\Gamma(a_{2}(1320)^{\pm} \to \pi^{\pm} \gamma)/
\Gamma(\rho^{\pm} \to \pi^{\pm} \gamma)=3.49$, 
which is independent from choice of the parameter $g_{M}$, 
in satisfactory agreement with data: $4.22\pm 0.876$. Thus it turns out that 
our form factor yields a desirable damping effect. 

On the other hand, concerning the $a_{1}(1260)$($b_{1}(1235)$) 
$\to\pi \gamma$ process, 
it seems that our results are 
poorly-fitted to the experiments \cite{Zielinski:1984,Collick:1984}, 
being both only measurements quoted in Ref. \cite{PDG:2012}. 
However, it was pointed out that in Ref.~{\cite{SELEX:2001}}, 
no evidence of the $a_{1}$ was found in the another 
charge-exchange photo-production experiment \cite{Condo:1993}, 
while a clear $a_{2}(1320)$ signal was observed. 
This results suggest that total width of $a_{1}(1260)$ meson is 
extremely large or radiative $\pi\gamma$ width is rather small. 
\footnote{Related works treating these 
mesons from another viewpoint 
have been reported \cite{Roca:2004, Nagahiro:2008, 
Nagahiro:2009, Molona:2011}. }
In any case, a high-statistics confirmation of these process by 
the COMPASS would be desirable. 

Concerning the $j^{\rm (convec.)}$ process, 
there are two possible ways for evaluating numerically 
the factor $M_{I}^2-M_{F}^2$ in the couplings $g_{b_{1}}$ and $g_{\pi_{2}}$: 
In Table \ref{tab:3}, we apply $M_{I}^2-M_{F}^2=N\Omega $ in the Model A by using 
Eq.(\ref{Eq:4}), while $M_{I}^2-M_{F}^2=2M_{I} |\bs{q}_{\gamma}|$ in the Model B.
In both cases, it is shown that 
the radiative decay widths of $\pi_{2}(1670)$ into $\pi \gamma$
have large fraction, predicted\footnote{
Here we use $\Gamma_{\rm tot}(\pi_{2}(1670))=260\pm 9~ {\rm MeV}$ 
taken from Ref.~\cite{PDG:2012}. 
} as 
${\rm Br}(\pi_{2}(1670)^{\pm} \to \pi^{\pm} \gamma)
=\Gamma (\pi_{2}(1670)^{\pm} \to \pi^{\pm} \gamma)/
\Gamma_{\rm tot}(\pi_{2}(1670))=(2.0 \pm 0.069)\times10^{-3} $ for the Model A, while 
$(1.3 \pm 0.045) \times10^{-3} $ for the Model B, respectively. 
Thus it must have been detected at the COMPASS Primakoff 
measurements with 
enough statistics. 

In summary we have investigated the radiative $\pi \gamma$ decays 
of the excited light-quark mesons in the COQM. 
Radiative decay widths of $D$-wave excited mesons are predicted. 
We expect that forthcoming experiments at COMPASS will make these predictions to verify.

%%%%%%%%%%%%%%%%%%%%%%%%%%%%%%%%%
%\ack
%Acknowledgements should go here.
%%%%%%%%%%%%%%%%%%%%%%%%%%%%%%%%%


\begin{thebibliography}{99}
\bibitem{COMPASS:2011} C. Adolph {\it et al}. (the COMPASS collaboration), \PRL{108,192001,2012}
\bibitem{COMPASS:2012-1}S. Grabm{\"u}ller, for the COMPASS collaboration, in Proceedings of the Sixth International Conference on Quarks and Nuclear Physics (QNP2012), Ecole Polytechnique Palaiseau, Paris, 2012. 
\bibitem{COMPASS:2012-2} F. Nerling, for the COMPASS collaboration, EPJ Web Conf. 37 (2012) 01016,
arXiv:1208.0487 [hep-ex]. 
\bibitem{Primakoff:1951}H. Primakoff, \PR{81,899,1951}
\bibitem{Pomeranchuk-Shmushkevich:1961} I. Ya. Pomeranchuk, I. M. Shmushkevich, 
\NP{23,452,1961}
\bibitem{Zielinski:1987} M. Zielinski, Acta Phys. Pol. B, {\bf18}, 455 (1987).
\bibitem{PDG:2012} J. Beringer {\it et al}. (Particle Data Group), \PRD{86,010001,2012}
\bibitem{Zielinski:1984} M. Zielinski {\it et al}., \PRL{52,1195,1984}
\bibitem{Collick:1984} B. Collick {\it et al}.,\PRL{53,2374,1984} 
\bibitem{Ishida-Yamada-Oda:1989} S. Ishida, K. Yamada, and M.Oda, 
\PRD{40,1497,1989} 
\bibitem{Ishida-Morikawa-Oda:1998} S. Ishida, A. Morikawa and M.Oda, 
\PTP{99,257,1989}
\bibitem{Yukawa:1953} H. Yukawa, \PR{91,415,1953} 
\bibitem{Takabayashi} T. Takabayasi, \NC{33,668,1964}\\
S. Ishida and J.Otokozawa, \PTP{47,2117,1972}\\
Y. S. Kim, M. E. Noz, \PRD{8,3521,1973}
\bibitem{Feynman-Kislinger-Ravndal} R. P. Feynman, M. Kislinger and 
F. Ravndal, \PRD{3,2706,1971}
\bibitem{Lipes:1972} R. G. Lipes, \PRD{5,2849,1972}\\
 S. Ishida and J. Otokozawa, \PTP{53,217,1975} 
\bibitem{Oda:1999} R. Oda, K. Yamada, S. Ishida, M. Sekiguchi and H. Wada, 
\PTP{102,297,1999}
\bibitem{Godfrey-Isgur:1985} S. Godfrey, N. Isgur,\PRD{32,189,1985}
\bibitem{Rosner:1981} J. L. Rosner, \PRD{23,1127,1981}
\bibitem{SELEX:2001} V.V. Molchanov,{\it et al}., \PLB{521,171,2001}
\bibitem{Condo:1993} G. Condo {\it et al}., \PRD{48,3045,1993}
\bibitem{Roca:2004} L. Roca, J. E. Palomar, and E. Oset, \PRD{70,094006,2004}
\bibitem{Nagahiro:2008} H. Nagahiro, L. Roca, E. Oset, \PRD{77,034017,2008}
\bibitem{Nagahiro:2009} H. Nagahiro, L. Roca, A. Hosaka, E. Oset, \PRD{79,014015,2009}
\bibitem{Molona:2011} R. Molina, H. Nagahiro, A. Hosaka, and E. Oset, \PRD{83,094030,2011}
\end{thebibliography}
\end{document}